\documentclass[reprint,prb,superscriptaddress]{revtex4-2} 
\usepackage{graphicx}
\graphicspath{{figs/}}

\usepackage[usenames,dvipsnames]{color}
\usepackage{soul}
\usepackage[colorlinks=true,
            linkcolor=Blue,
            citecolor=Blue,
            urlcolor=Blue,
            pdftitle={Repulsive disk potential},
            bookmarks=true
            ]{hyperref}

\usepackage[all]{hypcap}

\usepackage{amsmath}
\newcommand\w{\omega}
\newcommand\pb{\mathbf{p}}
\newcommand\qb{\ensuremath\mathbf{q}}

\def\blue{\color{black}}

\begin{document}

\title{\blue Pairing from repulsion in a two-dimensional Fermi gas with soft-core interactions}
%\title{On the pairing glue in repulsive Fermi gases with soft-core interactions}

\author{Ahmet Keles}
\affiliation{Department of Physics, Middle East Technical University, Ankara 06800, Turkey}
\author{Xiaopeng Li}
\affiliation{State Key Laboratory of Surface Physics, Key Laboratory of Micro and Nano Photonic Structures (MOE), and Department of Physics, Fudan University, Shanghai 200433, China} 
\affiliation{Shanghai Qi Zhi Institute, AI Tower, Xuhui District, Shanghai 200232, China}
\affiliation{Shanghai Research Center for Quantum Sciences, Shanghai 201315, China} 
\author{Erhai Zhao}
\affiliation{Department of Physics and Astronomy, George Mason University, Fairfax, Virginia 22030, USA}

\begin{abstract}
%Recent quantum gas experiments are ushering in an era of fine-tailored two-body 
%interatomic 
%interactions. 
%These include for example atoms with long-range dipole-dipole interactions and Rydberg-dressed atoms 
%created by optical coupling to highly excited electronic states. 
%
%These systems differ from the ``canonical system" of cold atoms with %hard-core 
%contact interactions characterized by a single parameter, the scattering length.
{\blue 
We investigate a model many-body system of 
spinless Fermi gas in two dimensions, where the
bare two-body interaction is repulsive and takes the form of a soft-core disk potential.
We obtain the zero temperature phase diagram of this model by 
numerical functional renormalization group (FRG), which 
%goes beyond leading order perturbation theory and 
retains the effective interaction vertices in all channels to provide a detailed picture of how 
Cooper pairing emerges %from pure repulsion 
under the renormalization flow.
The repulsion drives the system to a series of
%nontrivial 
superfluid states with higher angular momentum paring, for example in the
$f$- and $h$-wave channels instead of the $p$-wave channel. This is in sharp contrast
to the original Kohn-Luttinger mechanism where pairing of very large
angular momenta and exponentially small transition temperature was predicted.
We trace the stabilization and enhancement of $f$- and $h$-wave pairing 
back to the momentum dependence of the bare interaction.  
A perturbative calculation %which is justified in the dilute limit and 
is carried out to show that while the second order Kohn-Luttinger diagrams provide a qualitative understanding of the
onsets of the various superfluid phases, they are unable to accurately capture the phase boundaries predicted by FRG. 
Our findings suggest that tuning the shape of the interaction potential offers a promising route to achieve 
stronger ``pairing glue" and to realize nontrivial superfluid phases in repulsive Fermi gases beyond the scope of the original Kohn-Luttinger analysis.
}
\end{abstract}

\maketitle

\section{Introduction}
{\blue It remains a long-standing goal to realize  non-$s$-wave superfluids in ultracold Fermi gases.
For example, the conventional wisdom to realize the $p_x +ip_y$ state}
in spin-polarized (single-species) Fermi gases is to bring it close to a $p$-wave Feshbach
resonance where the $p$-wave interaction between two fermionic atoms becomes attractive 
\cite{PhysRevLett.90.053201,PhysRevA.70.030702,PhysRevLett.95.230401,PhysRevA.96.062704}. 
Unfortunately, this effort has been hampered by severe three-body losses near the resonance. 
Despite the recent success in improving the gas lifetime in one- and three-dimensional optical lattices 
\cite{PhysRevLett.125.263402,marcum2020suppression,luciuk2016evidence,venu2023unitary},
it remains a challenge to suppress the atom loss.
Here in this paper, we explore an alternative route that does not require $p$- or higher-wave resonances.
In particular, we address the following questions:
Is there room for superfluidity in polarized Fermi gases if the bare interaction is {\it purely repulsive}? 
{\blue If so, in which parameter regimes is the superfluid transition temperature $T_c$ enhanced and thus
more accessible by
experiments?}

Central to these questions is the issue of ``pairing glue" in a repulsive Fermi liquid. 
The term ``pairing glue" is often used in the literature on quantum materials, and it refers 
to the microscopic mechanism that binds the fermions into Cooper pairs \cite{anderson-glue,monthoux2007superconductivity}.
%i.e., how the effective attraction is induced. 
It is well known that in most conventional $s$-wave superconductors, phonons act as the glue \cite{degennes2018superconductivity}.
On the other hand, while there is no consensus yet, spin fluctuations are likely responsible for the $d$-wave pairs observed in 
cuprate superconductors, or the 
the putative $d$-wave superfluid phase of the repulsive Fermi-Hubbard model (for a review see for instance \cite{maiti2013superconductivity,Kagan_2015}).
%(There is yet a consensus on this issue.) 
For continuum gases of spinless fermions, however, neither lattice vibration nor spin fluctuation is 
present, so only density fluctuation can step in to make the glue.
Since the system cannot remain a Fermi liquid down to zero temperature,
it is long believed that many-body effects will turn the 
repulsive bare interaction into attractive effective interaction in certain pairing channels.
In other words, the force between two fermions may flip sign under renormalization \cite{shankar1994}.

%% Kohn luttinger background
A well-known example is the Kohn-Luttinger (KL) mechanism discovered by
W. Kohn and J. M. Luttinger back in 1965 \cite{PhysRevLett.15.524,PhysRev.150.202}. % in a paper titled ``New Mechanism for Superconductivity." 
They showed that for spin $1/2$ fermions in three dimensions (3D) with weak short-range 
repulsive interactions, the effective interaction $\Gamma_\ell$ in the $\ell$-th angular momentum channel 
always turns attractive for sufficiently high partial waves, e.g. some odd $\ell \gg 1$. 
This means that a repulsive Fermi liquid is always unstable against pairing (there may be other
competing instabilities as well), albeit
the corresponding $T_c\propto e^{-\alpha \ell^4}$, with $\alpha$
some constant, is exponentially small \cite{PhysRevLett.15.524}. Fay and Layzer generalized the KL analysis for large $\ell$
to include small $\ell$. In the dilute limit, regardless the strength of the interactions, they
found the dominant instability is toward a $p$-wave superfluid with $\ell = 1$ \cite{PhysRevLett.20.187}.
Kagan and Chubukov reached the same conclusion and computed the $T_c$ \cite{kagan1988possibility}.
The KL effect in two dimensions (2D) requires a more delicate analysis beyond second order perturbation theory,
but as shown by Chubukov \cite{PhysRevB.48.1097}, the dominant pairing instability remains in
the $p$-wave channel. We stress that all these results were obtained for spin-$1/2$ {fermions}
%Fermi gases,
where the bare interaction is replaced by a pseudopotential that can be parametrized by the $s$-wave scattering length $a$ \cite{PhysRevLett.15.524}. 

\begin{figure}
\includegraphics[width=0.48\textwidth]{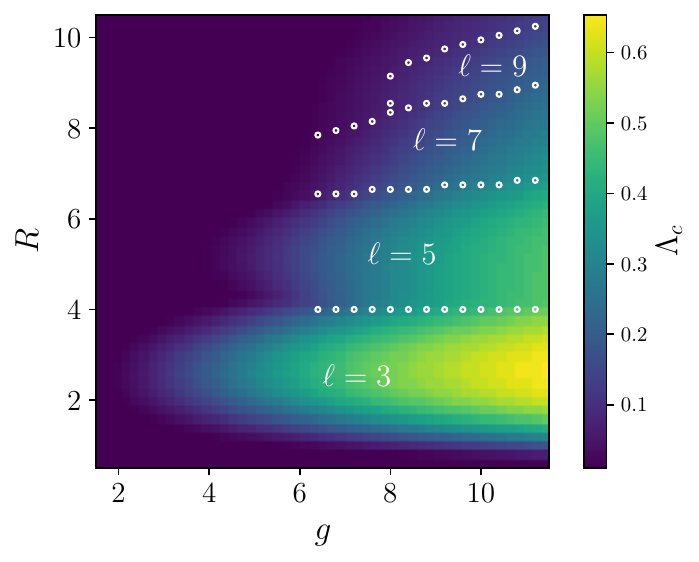}
\caption{The FRG phase diagram of a repulsive Fermi gas in 2D: the superfluid phases resemble four fingers of 
a hand (with the thumb $\ell =1$ missing). The Cooper pair angular momentum is $\ell=3$ ($f$-wave), $5$ ($h$-wave), $7$ and $9$
respectively. The model and the dimensionless interaction strength $g$ are defined in Section \ref{sec:model}, and
the interaction range $R$ is measured in $1/k_F$, the inverse Fermi momentum. The empty circles mark the phase boundaries,
and the ``critical scale" $\Lambda_c$
in false color gives a rough estimate of the superfluid $T_c$.}
\label{fig:phase-diagram}
\end{figure}

The {\blue main goal} of this paper is to understand how repulsion drives pairing in spinless Fermi gas
in two dimensions {\blue from the modern perspective of functional renormalization group (FRG). 
We focus on a simple model of bare interaction in the form of
the disk potential Eq. \eqref{disk} which is often referred to as the 
square (or step function) potential in quantum mechanics textbooks. 
%In the limit of large $V_0$, it becomes the hard-disk potential
In order to go beyond the aforementioned KL analysis, 
we treat the many-body problem using FRG
which goes beyond leading order perturbation theory and retains the interaction vertices in all (e.g. pairing,
density wave, and Pomeranchuk) channels. We solve the FRG flow equation numerically
to obtain the full phase diagram and compare the transition temperature in
different parameter regions}. The main results are summarized in Fig.~\ref{fig:phase-diagram}.
{\blue We observe that, surprisingly, the behaviors of this system differ significantly from the
%will show below that spin-polarized Fermi gases in 2D with a disk  two-body interaction potential 
classical KL results outlined in the preceding paragraph. 
For instance, $f$-wave or $h$-wave (instead of $p$-wave) superfluid states are stablized,
and their transition temperatures are not exponentially small.}

{\blue To gain further understanding of the numerical FRG result}, 
we also carry out a perturbative calculation which becomes accurate in the dilute
(low density) limit. We show that evaluation of the so-called KL diagrams for our model
yields results in qualitative agreement with FRG. 
The perturbative calculation enables us to see how and when the effective interaction $\Gamma_\ell$
turns negative, and why it differs significantly from the well-known KL physics in spin-1/2 systems.
The calculation also illustrates the limitations of perturbation theory. For example, 
the predicted phase boundaries (see Fig. \ref{fig:perturbation-boundary}) 
deviate significantly from the FRG phase diagram (Fig. \ref{fig:phase-diagram}) 
which is much more accurate because it includes many-body processes  well beyond
the KL diagrams. 

%A critic might dismiss the whole calculation by categorizing our choice of the 
{\blue The bare potential Eq. \eqref{disk}, as a simple model to elucidate the 
intriguing many-body physics, may not be easily realized in experiments. Our model choice however is not arbitrary and 
in fact inspired by}
the interaction potential in Rydberg-dressed Fermi gases which recently became available
in experiments \cite{PhysRevX.11.021036}.  In Ref. \cite{PhysRevA.101.023624}, three of us discovered that an $f$-wave superfluid
naturally emerges in these systems even when the bare Rydberg-dressed interaction is repulsive.
The disk potential here retains the soft-core part of Rydberg-dressed potential
but discards its long-range tail. {\blue By comparing the phase diagrams of the two models, one makes an important observation: 
it is the repulsive core, rather than the long-range tail, that is crucial to $f$-wave pairing. 
The current model also features a much richer phase diagram.
The perturbative analysis (Section IV) and its comparison 
against FRG are also new results
%For these reasons, the calculations and analysis presented here 
% that significantly exceeds 
beyond the scope of Ref. [21].}
%
%should not be taken as 
%a repeat of those in Ref. \cite{PhysRevA.101.023624}.}
%{Of course, the long-range tail does play a role in determining  the detailed features such as the phase boundaries \cite{PhysRevA.101.023624}.}  

%Besides its connection to Rydberg-dressed quantum gases, model Eq.~\eqref{disk} was chosen to 
%show that 
{\blue Our results for model Eq.~\eqref{disk} makes it clear that}
{\it the shape of the bare interaction matters} to enhancing the pairing
glue in repulsive Fermi gases. A nice feature of the disk-potential $V(r)$ is that
its Fourier transform $v({q})$ develops oscillations and becomes attractive for 
certain range of momenta, see Eq. \eqref{vq}. This is in contrast to previous works on spin-$1/2$ Fermi gases, 
where $v({q})$ is usually assumed to be a constant $u$. Under renormalization, these attractive 
segments of $v({q})$ feed to the flow of the effective interaction $\Gamma_\ell$
toward negative values, 
eventually leading to a slew of superfluid phases with $\ell =3, 5, 7, 9...$ in Fig. \ref{fig:phase-diagram} {\blue 
(this feature is absent in Ref. \cite{PhysRevA.101.023624})}. 
In the original KL picture \cite{PhysRevLett.15.524}, the effective interaction between two fermions acquires 
a long-range oscillatory part because of the sharp Fermi surface, which is related to the
Friedel oscillations in real space. In our case, we have not only a sharp Fermi surface 
(a step function in momentum space), but also a sharp two-body interaction potential (a step function
in real space). This {\it double whammy} also partly explains why the of $T_c$ of
these superfluid phases is not exponentially small as in the original KL analysis. 
{\blue To our knowledge, the importance of the interaction shape has not received a lot of attention
in the literature.
We hope the model study presented here can stimulate new ideas
to engineer stronger pairing glues by shaping the bare interactions. Our results suggest
that this is a promising route to observe higher angular momentum
{pairing} in repulsive Fermi gases.}

\section{Model and bare interaction}
\label{sec:model}
Our model is a spin-polarized (spinless) Fermi gas in 2D with the short-range interaction potential
\begin{equation}
V(r)=V_0\theta(R-r). \label{disk}
\end{equation}
Here $r$ is the distance between two fermions, $\theta(x)$ is the Heaviside step function,
$R$ is the radius of the disk, and $V_0>0$ is the interaction strength. In the limit of large $V_0$, 
$V(r)$ gives the hard-disk potential (not hard-sphere because we are in 2D). For this reason, we shall call Eq. \eqref{disk}
the soft-core disk potential. Let $k_F$ be the Fermi momentum, $m$ the mass of the fermion, then
the density of state is $\mathcal{N}=m/2\pi \hbar^2$ (we will set $\hbar=1$ hereafter). We define dimensionless parameter
\begin{equation}
g = (2\pi R^2 V_0) \mathcal{N} 
\end{equation}
which measures the strength of interaction. Another independent dimensionless parameter
is $k_FR$, which measures the range of the interaction in units of $1/k_F$. In Ref. \cite{PhysRevA.101.023624},
similar parameters were defined for the Rydberg-dressed Fermi gases.

\begin{figure}
\includegraphics[width=0.48\textwidth]{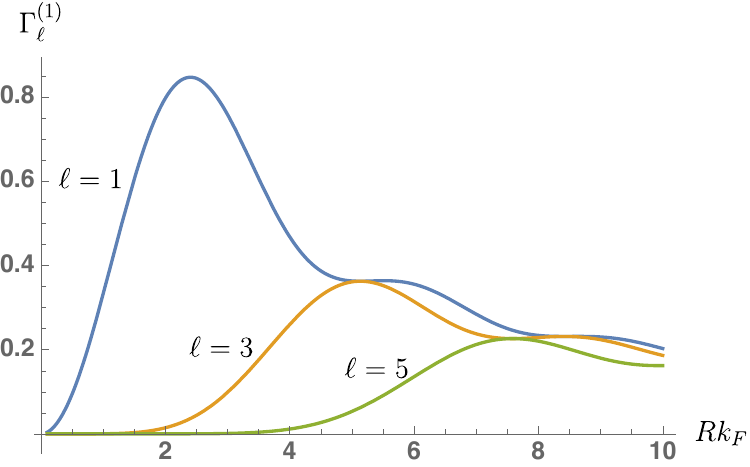}
\caption{The bare interaction $\Gamma^{(1)}_\ell$ in angular momentum channel $\ell=1$ (blue), $3$ (orange), and $5$ (green),
all being repulsive. 
$\Gamma^{(1)}_\ell$ is defined in Eq. \eqref{decompose} for two fermions on the Fermi surface and measured in unit of $(2\pi R^2V_0)$.}
\label{fig:bare}
\end{figure}

Let us look at the bare interaction in more detail. 
The Fourier transform of $V(r)$ is given by the Bessel function 
\begin{equation}
v(q) = 2\pi V_0R^2 \frac{J_1(qR)}{qR} \label{vq}
\end{equation}
with $q=|\mathbf{q}|$, and $\mathbf{q}$ is the momentum. The function
$v(q)$ is a damped oscillation and turns negative repeatedly.
For example, its first negative minimum is at $x=qR=5.136$,
where $J_1(x)/x=-0.06614$. Again, it is useful to compare it to the
{Meijer G-function discussed in Ref.~\onlinecite{PhysRevA.101.023624}} 
which only has one negative minimum.
For pairing, the relevant bare interaction is between two fermions on the Fermi surface at momentum $\mathbf{p}_F$ and $\mathbf{p}'_F$ respectively, 
i.e. $v(|\mathbf{p}_F-\mathbf{p}'_F|)$. Let $\phi$ be the angle between $\mathbf{p}_F$ and $\mathbf{p}'_F$, the bare interaction can be written as 
\begin{equation}
\Gamma^{(1)}(\phi) = v(2k_F|\sin \frac{\phi}{2}|). \label{bareint}
\end{equation}
Here the superscript of $\Gamma^{(1)}$ emphasizes that this is the leading order (to order $V_0$) contribution
from the perspective of perturbation theory. We can decompose $\Gamma^{(1)}(\phi)$ into angular momentum channels
by defining
\begin{equation}
\Gamma^{(1)}_\ell = \int d\phi \Gamma^{(1)} (\phi) \cos (\ell \phi),\;\;\; \ell=1,3,5,...
\label{decompose}
\end{equation}
Only odd $\ell$ values are taken, because we are dealing with spinless fermions.
The integral in Eq. \eqref{decompose} can be evaluated analytically by exploiting the properties of Bessel functions.
The results are plotted in Fig. \ref{fig:bare} for $\ell=1,3,5$.
We observe that $\Gamma^{(1)}_\ell$ are all positive.
{This is as expected, for the bare repulsion does not directly lead to Cooper pairing (to order $V_0$).}
%This is exactly what one expects: the bare repulsive interaction does not directly lead to pairing (to order $V_0$). 
We need many-body effects to induce effective attraction to overcome the bare repulsion.

\section{Main results from FRG}

We analyze the interacting fermion problem by Functional Renormalization Group (FRG)
\cite{RevModPhys.84.299,peterkopietz2010}.
Technical details of the FRG approach to 2D continuum Fermi gases can be found in  
Refs.~ \onlinecite{PhysRevA.101.023624}
and \onlinecite{PhysRevA.94.033616}, 
and our implementation here follows Ref.~\onlinecite{PhysRevA.101.023624} closely. 
For examples of FRG applied to Fermi gases on optical lattices, see Refs.~\cite{PhysRevLett.108.145301,
PhysRevA.87.043604,PhysRevLett.110.155301}.
We obtain the zero temperature phase diagram using the following procedure.
Starting from an ultraviolet scale $\Lambda_{UV}$, where the effective interaction equals to the anti-symmetrized
bare interaction, we slowly slide down the momentum scale $\Lambda\rightarrow \Lambda-\delta \Lambda$ by successively 
integrating out the higher energy, shorter wavelength fluctuations. The result is 
a set of coupled flow equations, e.g., for the self-energy $\Sigma$ 
\begin{equation}
  \partial_\Lambda \Sigma_{1',1} = -\sum_{2} S_{2} \Gamma_{1',2;1,2},
  \label{eq:sigma}
\end{equation}  
and for the four-fermion vertex $\Gamma$
\begin{align}
    \partial_\Lambda\Gamma_{1',2';1,2}& = \sum_{3,4} \big( G_{3} S_{4} + S_{3} G_{4}\big)
    \big[
      \frac{1}{2} \Gamma_{1',2';3,4} \Gamma_{3,4;1,2} \nonumber \\
      &-\Gamma_{1',4;1,3}\Gamma_{3,2';4,2} 
     +\Gamma_{2',4;1,3}\Gamma_{3,1';4,2} 
    \big].
    \label{eq:gamma}
\end{align}  
Here 
%to clarify the overall structure of the equations, 
$1,2$ ($1',2'$) labell the incoming (outgoing) legs of the effective interaction $\Gamma$,
and we have used the short-hand notation $1\equiv(\omega_1,\mathbf{p}_1)$ to denote
the fermion frequency $\omega$ and momentum $\mathbf{p}$. 
The sum in Eqs. \eqref{eq:sigma}-\eqref{eq:gamma} includes integration over frequency 
and momentum, e.g., 
\[
\sum\limits_3 (...)= \int \frac{d\w_3 d^2\pb_3}{(2\pi)^3} (...).
\]
Eq. \eqref{eq:sigma} and Eq. \eqref{eq:gamma} can be represented diagrammatically.
The first term inside the square bracket in Eq. \eqref{eq:gamma} gives
the BCS diagram in the
particle-particle channel, while the second (third) term gives the 
ZS (ZS') diagram in the particle-hole channel. 
Here ZS stands for zero sound~\cite{RevModPhys.66.129}.  
The term $(G_{3} S_{4} + S_{3} G_{4})$ on the right hand side of Eq. \eqref{eq:gamma} 
is the analogue of 
{the polarization bubble, but it has a crucial difference as it involves two
scale-dependent Green functions defined by} 
\begin{align}
    G_{\w,\pb} =
    \frac{\theta(|\xi_\pb|-\Lambda) }{i\w-\xi_\pb-\Sigma_{\w,\pb}  } ,\;\;
    S_{\w,\pb} =
    \frac{\delta(|\xi_\pb|-\Lambda) }{i\w-\xi_\pb-\Sigma_{\w,\pb}  },
    \label{eq:propagators}
\end{align}
where $\xi_\pb = \pb^2/2m-E_F$, with $E_F$ the Fermi energy. We stress that $G$, $S$, $\Sigma$ and $\Gamma$ all depend on the sliding scale
$\Lambda$, even though we have suppressed the $\Lambda$-dependence in our notation for brevity. 

The FRG flow equations are formally exact, but in practice they must be 
truncated and approximated in order for the numerical calculation to become feasible.
Higher order contributions have been dropped from Eqs. \eqref{eq:sigma} and 
\eqref{eq:gamma}. We further neglect the frequency
dependence of $\Gamma$ and drop $\Sigma$, which is typically 
not necessary to reveal the leading instabilities. Finally, we project the 
momenta radially onto the Fermi surface because the angular dependence is most relevant,
and accordingly we discretize the Fermi surface evenly into $N$ patches. 
Then, $\Gamma$ is reduced  to a three-dimensional array,
\[
\Gamma_{1',2';1,2}\rightarrow\Gamma(\pb_{F1}',\pb_{F2}',\pb_{F1})\rightarrow \Gamma_{i,j,k}.
\]
Here only three momentum variables are needed thanks to the conservation of the total momentum,
and $i,j,k=1,2, ..., N$ are the patch indices giving the angular position on the circular Fermi surface. 
{\blue
We stress that similar truncation and approximation schemes have been extensively employed and 
benchmarked in the application of FRG to correlated electrons. For a detailed assessment and justification of these
steps, the readers may consult the review Ref. \cite{RevModPhys.84.299}. 
In principle, one can systematically include higher order
diagrams and take into account the frequency dependences. These improvements however come
with a steep increase in the requirement of computing resources.
}

Even with these simplifications, 
the computation remains heavy. For example, for an angular grid with $N=128$,
$\Gamma$ contains $N^3$, roughly 2 million, elements. We call them running couplings,
because they undergo nontrivial evolutions as $\Lambda$ is reduced. 
Among all the running couplings, the largest absolute value is denoted as
\[
\Gamma_{max}=\mathrm{max}|\Gamma_{i,j,k}|.
\] 
From $\Gamma$, we also construct the channel matrix for BCS pairing 
\begin{equation}
V_\mathrm{BCS}(\pb',\pb)=\Gamma(\pb',-\pb',\pb).
\end{equation}
and the channel matrix for charge density wave (CDW) order with wavevector $\qb$
\begin{equation}
V^\qb_\mathrm{CDW}(\pb',\pb)=\Gamma(\pb+\qb/2,\pb'-\qb/2,\pb-\qb/2) .
\end{equation}
Another example
is the Pomeranchuck channel
\begin{equation}
V_\mathrm{POM}(\pb',\pb)=\Gamma(\pb,\pb',\pb),
\end{equation}
%This corresponds to the forward scattering limit, 
the instability of which %in this channel
points to spontaneous deformation of the Fermi surface.
%These quantities will become important
%to determining the phase diagram below.
%
With these approximations, the flow equation \eqref{eq:gamma}
is solved numerically by sliding $\Lambda$ on a logarithmic grid from the ultraviolet (UV) scale 
$\Lambda_{UV}=E_F$ down to a very small infrared (IR) scale, e.g., $\Lambda_{IR}=0.01E_F$. 
Typically we have hundreds of grid points along the $\Lambda$ axis, %(the RG scale),
and at each RG step, the most time consuming part is the summation 
over internal lines, $\sum_{3,4}$ in Eq. \eqref{eq:gamma}.
The calculation is checked to ensure the result 
does not change upon further refining the angular or $\Lambda$ grids.

To detect possible many-body instabilities of the interacting Fermi gas, we monitor the flow of $\Gamma$ 
and look for signs of divergence as $\Lambda\rightarrow 0$. 
For example, a clear signal of divergence is when $\Gamma_{max}$ quickly exceeds a large threshold 
such as $100E_F$ at some ``critical value" $\Lambda = \Lambda_c$. 
%A larger value of $\Lambda_c$ indicates stronger divergence. 
In such cases we record $\Lambda_c$ and use it 
as an estimate of the $T_c$ of the corresponding broken symmetry phase.
In other cases (see Fig. \ref{fig:no-div} below), the flow continues smoothly down to 
$\Lambda_{IR}$, indicating the Fermi liquid is stable down to this temperature scale, within
the approximation and numerical precision of our calculation.
The channel matrices defined above provide a systematic way to identify the broken symmetry phases.
In each channel $ch\in\{\mathrm{BCS, CDW, POM}, ...\}$ and at each RG step,
we diagonalize the channel matrix $V_{ch}$ and record
its most negative eigenvalue $\Gamma_{min}^{ch}$ 
 (for density
waves, we also vary $\mathbf{q}$ to seek the lowest eigenvalue among all $\mathbf{q}$). 
The leading divergence can be easily identified by comparing all $\Gamma_{min}^{ch}$
as $\Lambda$ is reduced. The eigenvector of the most divergent $\Gamma_{min}^{ch}$ reveals the orbital symmetry
%e.g. $p$- or $f$-wave, 
of the incipient order. 

\begin{figure}
\includegraphics[width=0.48\textwidth]{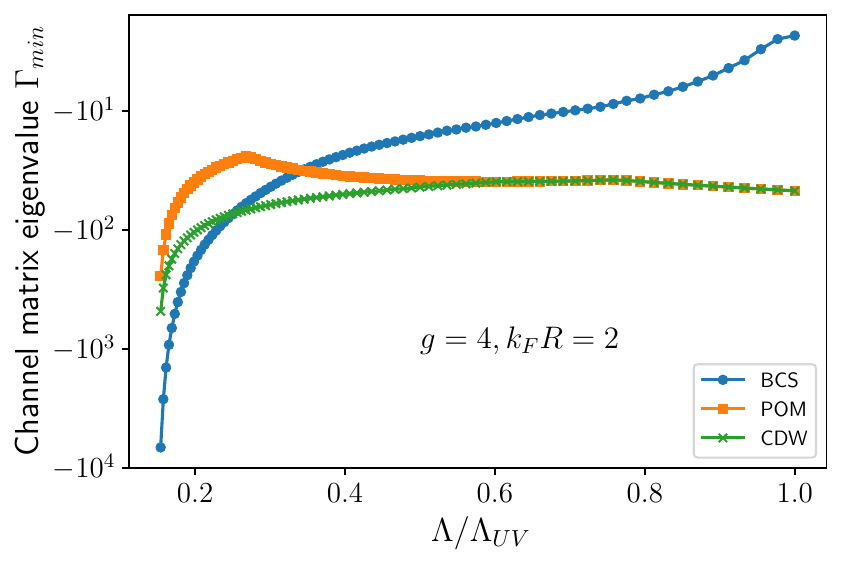}
\includegraphics[width=0.35\textwidth]{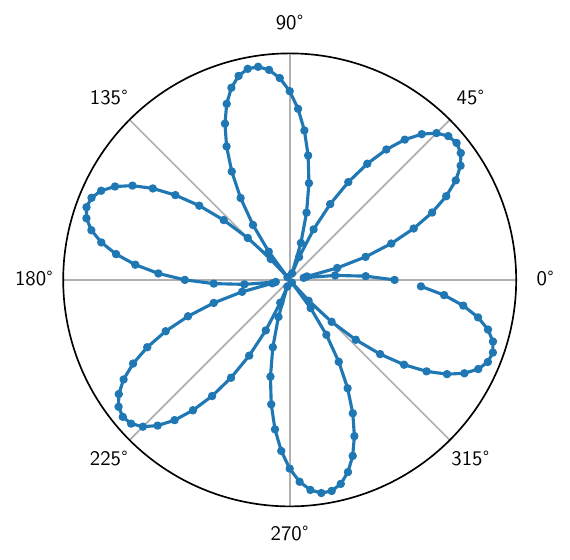}
\caption{The FRG flow for parameter $g=4$ and $k_FR=2$.
Upper panel: the most negative eigenvalue, $\Gamma_{min}$,  in the BCS, Pomeranchuck, and CDW channel.
As the sliding RG scale $\Lambda$ is reduced from $\Lambda_{UV}$, BCS becomes the leading instability.
Lower panel: the eigenvector $f_\mathrm{BCS}(\phi)$ corresponding to the BCS instability, with $\phi$ 
going from $0^\circ$ to $360^\circ$ around the Fermi surface. Its nodal structure
shows $f$-wave pairing with angular momentum $\ell=3$ (see main text).}
\label{fig:comp1}
\end{figure}

\begin{figure}
\includegraphics[width=0.48\textwidth]{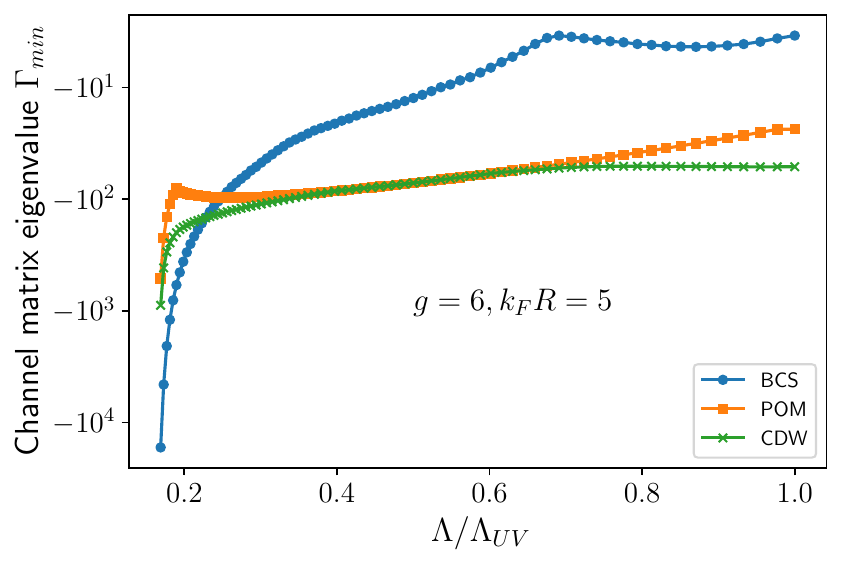}
\includegraphics[width=0.35\textwidth]{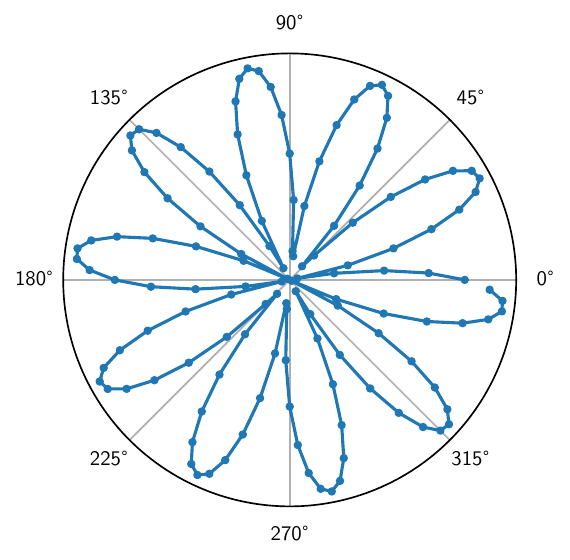}
\caption{
Evidence of $h$-wave pairing from FRG for parameter $g=6$ and $k_FR=5$.
Upper panel: the competition between the BCS, Pomeranchuck, and CDW instability.
Lower panel: the eigenvector $f_\mathrm{BCS}(\phi)$ can be fit by $A\cos (5\phi-\phi_0)$,
pointing clearly to $h$-wave pairing with $\ell=5$.}
\label{fig:comp2}
\end{figure}

Fig. \ref{fig:comp1} shows the competition of the BCS, CDW, and Pomeranchuck channel
for interaction strength $g=4$ and interaction range $k_FR=2$. We observe from the upper panel that long before $\Lambda_{IR}$ is reached, the 
BCS channel (in blue) develops into the leading divergence, with the other two channels trailing behind.
The polar plot in the lower panel shows the eigenvector for $\Gamma_{min}^\mathrm{BCS}$ 
% in the BCS channel 
as a function of $\phi$ %(introduced above Eq. 4) 
as it  varies from 0 to $2\pi$ around the Fermi surface.
It features 6 nodes, and can be fit nicely by $f_\mathrm{BCS}(\phi)=A\cos(3\phi-\phi_0)$.
The evidence unambiguously points to an $\ell=3$, or $f$-wave, superfluid phase.
Another example is shown in Fig. \ref{fig:comp2} for $g=6$ and $k_FR=5$. While the flow looks rather similar to Fig. 2 and
the leading instability remains in the BCS channel (upper panel), the eigenvector (lower panel) tells a different story. 
The orbital symmetry in this case is clearly different, suggesting an $\ell=5$, or $h$-wave superfluid instead. Yet another
example is shown in Fig. \ref{fig:no-div}. Here none of the channel matrix eigenvalues develops divergence
as $\Lambda_{IR}$ is reached. 

\begin{figure}
\includegraphics[width=0.48\textwidth]{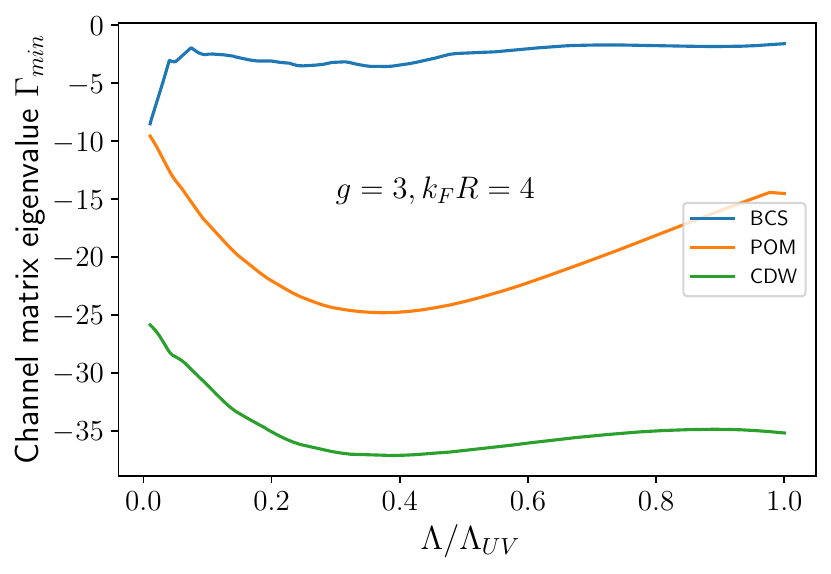}
\caption{Absence of long-range order for parameter $g=3$ and $k_FR=4$.
In contrast to Fig. \ref{fig:comp1} and Fig. \ref{fig:comp2}, no divergence is visible
as $\Lambda$ is reduced down to $\Lambda_{IR}$. }
\label{fig:no-div}
\end{figure}

Similar FRG analysis can be performed for other parameter values 
on the $(g,R)$ plane, and the results are summarized in the phase diagram shown in Fig. \ref{fig:phase-diagram}. 
The most striking feature of the phase diagram is a series of
superfluid phases with Cooper pair angular momentum $\ell = 3, 5,7,9$.
The empty circles mark the phase boundary, and the 
background false color shows the critical scale
$\Lambda_c$ serving as a rough estimate of the $T_c$ of each ordered phase.
The overall shape of the phase diagram resembles a hand with the index, middle, ring, and little finger. 
Note that $p$-wave superfluid with $\ell=1$, or the ``thumb,'' is missing. 

\begin{figure}
\includegraphics[width=0.48\textwidth]{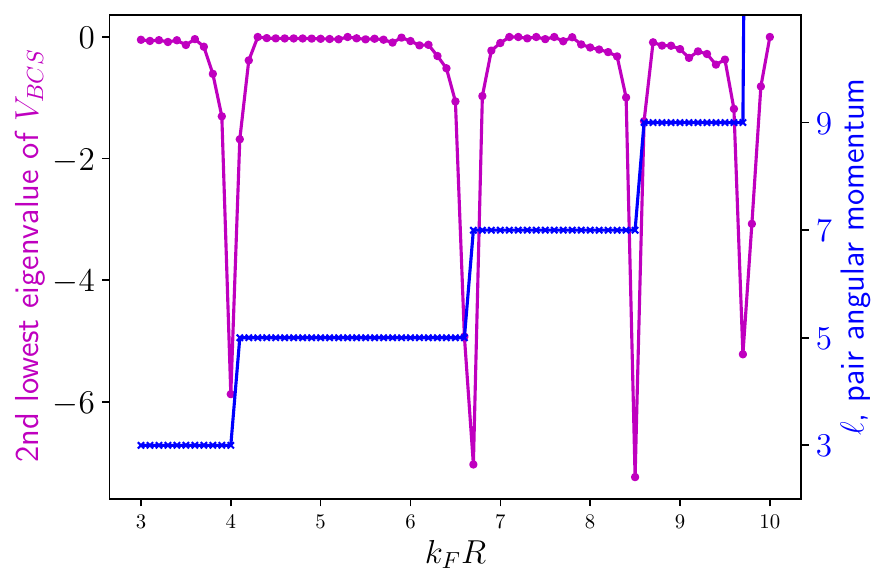}
\caption{Detecting the phase transitions along the vertical cut $g=9$
using two independent measures: (a) 
the pair angular momentum $\ell$ extracted from $f_\mathrm{BCS}(\phi)$, in color blue;
and (b) the second lowest eigenvalue of the BCS channel matrix $V_\mathrm{BCS}$ (rescaled by $10^3$), in color magenta.
The phase boundaries from the two methods agree with each other.}
\label{fig:phase-boundary}
\end{figure}

The phase boundaries (empty circles) in Fig.  \ref{fig:phase-diagram} are determined numerically as follows. 
In the first method, we decompose the eigenvector $f_\mathrm{BCS}(\phi)$ corresponding to 
$\Gamma_{min}^\mathrm{BCS}$ %(the most negative eigenvalue in the BCS channel) 
in the basis $\{\cos (\ell \phi)\}$ with odd $\ell\geq 1$.
%This can be done conveniently by carrying out a fast Fourier transform. 
%which is equivalent to counting the number of nodes. 
We find that there is only one dominant
$\ell$ component in each superfluid phase, and the value of $\ell$ jumps at the phase boundaries to form
 a terrace as $k_FR$ is varies along a vertical cut at constant $g=9$ (in blue, Fig. \ref{fig:phase-boundary}).
In the next method, we plot the second most negative eigenvalue of $V_{BCS}$ (in magenta, Fig. \ref{fig:phase-boundary}).
The idea is that as a phase boundary is approached, say going from the $f$-wave to the $h$-wave phase, 
the lowest two eigenvalues of $V_{BCS}$ are expected to become degenerate. Thus, 
the second lowest eigenvalue will take a dip whenever a phase boundary is crossed.
%providing an additional evidence for the phase transition.
Fig. \ref{fig:phase-boundary} shows that the phase boundaries
determined from these two independent measures agree well with each other. 
{\blue And there is no indication of phase coexistence.}

We stress that the empty circles in Fig. \ref{fig:phase-diagram} represent only part of the phase boundaries.
For small $g$ or large $R$,
the critical scale $\Lambda_c$ is pushed down toward $\Lambda_{IR}$, making it challenging to reliably 
determine the phase boundary using the methods outlined above. For this reason,
only well resolved data points are presented. For example, both the $f$- and 
$h$-wave superfluid persist to lower $g$ values with significantly reduced $T_c$,
and their phase boundaries are expected to extend to the left as well. Superfluid
phases with $\ell>9$ may exist at larger $g$ and $R$ values (not shown in Fig. \ref{fig:phase-diagram}), they are not well resolved
due to the limitation of our angular grid and the diminishing $T_c$ values.

Despite the apparent simplicity of our model, the phase diagram in Fig. \ref{fig:phase-diagram}
is quite rich and to our knowledge 
has not been reported before. Let us recall that generalizing the Kohn-Luttinger analysis to spin-$1/2$ Fermi gas with short-range
repulsion in 2D predicts a $p$-wave superfluid state \cite{PhysRevB.48.1097}, which has gone missing in our case. It is also worthwhile to compare Fig. \ref{fig:phase-diagram} 
to the phase diagram of the Rydberg-dressed Fermi gas in 2D, which harbors an $f$-wave superfluid
that becomes intertwined with, and eventually yields to a CDW as the interaction range is increased
\cite{PhysRevA.101.023624}.  
Here, we do not see a CDW phase, because it is pushed to very high $g$ values, $g>15$, 
according to the random phase approximation. 
Instead, we see the emergence of a series of superfluid phases with higher angular momentum pairing.

Since the FRG calculation involves delicate interplays of particle-particle and particle-hole fluctuations 
on a sliding momentum/energy scale, one might wish a simpler ``explanation" of how the bare repulsion 
is turned into a pairing glue. In the next section, we shed more light on these phases using perturbation theory.

\section{Insights from perturbation theory}

The superfluid phases occupy a large portion of the parameter space in Fig.~\ref{fig:phase-diagram}. A perturbation expansion
in power series of $V_0$ will not be justified everywhere, e.g. when $V_0$ or  $g$ is large. However, it is well recognized that in the
dilute, low-density limit (corresponding to small $k_FR$), a perturbative expansion is possible even for large $g$ \cite{baranov1992superconductivity}. With
these caveats in mind, our main objective in this section is to look for the trends (rather than the exact numbers) suggested by perturbation theory.

\begin{figure}
\includegraphics[width=0.48\textwidth]{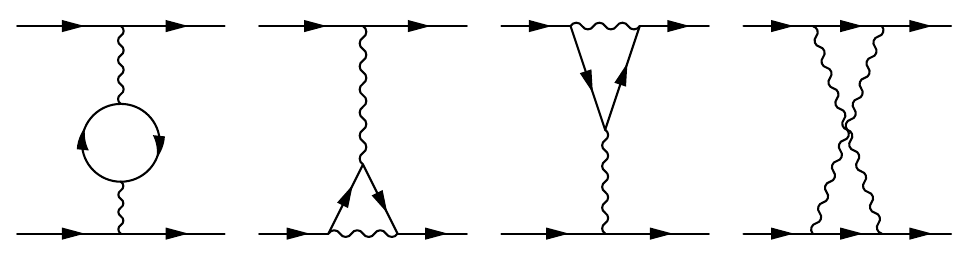}
\caption{The four Kohn-Luttinger diagrams as 
the second order contributions to the effective interaction.
From left to right are $D_1$, $D_2$, $D_3$ and $D_4$.
}
\label{fig:KLdiagrams}
\end{figure}

To the first order of $V_0$, the effective interaction is given by the bare repulsion in Eq. \eqref{bareint}.
%Despite the fact that $v(\mathbf{q})$ goes negative in certain regions, 
As shown in Fig. \ref{fig:bare}, all its angular components $\Gamma^{(1)}_\ell$ 
are positive. For $k_FR\leq 2$, $\Gamma^{(1)}_{\ell=1}$ dominates, with all other components negligibly small. To order $V_0^2$, the corrections to the effective interaction in the
Cooper channel consist of four contributions shown in Fig. \ref{fig:KLdiagrams}, often referred to as the Kohn-Luttinger diagrams \cite{PhysRevLett.15.524}.  They are vertex functions describing a fermion pair ($\mathbf{p}$,$-\mathbf{p}$) being scattered to ($\mathbf{p}'$,$-\mathbf{p}'$) that involves two bare interactions (wavy lines)
and two internal fermion propagators (solid lines). 
The first diagram contains a particle-hole bubble,
\[
D_1 = -i\int \frac{d^2\mathbf{k}d\omega}{(2\pi)^3} v(\mathbf{q})v(\mathbf{q})G_0(k)G_0(k+q).
\]
Here for spin-polarized fermions, the factor 2 associated with the fermion bubble is absent but the negative sign is retained. $G_0$ is the bare fermion Green function at $T=0$
(different from the scale-dependent Green function in the previous section), the 4-momentum $k=(\omega,\mathbf{k})$, and similarly $q=(\Omega,\mathbf{q})$ with
the momentum transfer
$\mathbf{q} = \mathbf{p}' - \mathbf{p}$.
% and the total momentum $\mathbf{Q} = \mathbf{p}' + \mathbf{p}$.
The second diagram contains the vertex correction
\[
D_2 = i\int \frac{d^2\mathbf{k}d\omega}{(2\pi)^3} v(\mathbf{q}) v(-\mathbf{p}-\mathbf{k})G_0(k)G_0(k+q).
\]
The third diagram is very similar to the second,
\[
D_3 = i\int \frac{d^2\mathbf{k}d\omega}{(2\pi)^3} v(\mathbf{q}) v(\mathbf{p}'-\mathbf{k})G_0(k)G_0(k+q).
\]
And the fourth diagram is the exchange scattering
\[
D_4 = i\int \frac{d^2\mathbf{k}d\omega}{(2\pi)^3} v(\mathbf{p}-\mathbf{k}) v(\mathbf{p}'-\mathbf{k})G_0(k)G_0(k-p-p').
\]
To evaluate these diagrams, first the $\omega$ integral is carried out analytically, then 
the integration over $\mathbf{k}$ is computed numerically.

\begin{figure}
\includegraphics[width=0.45\textwidth]{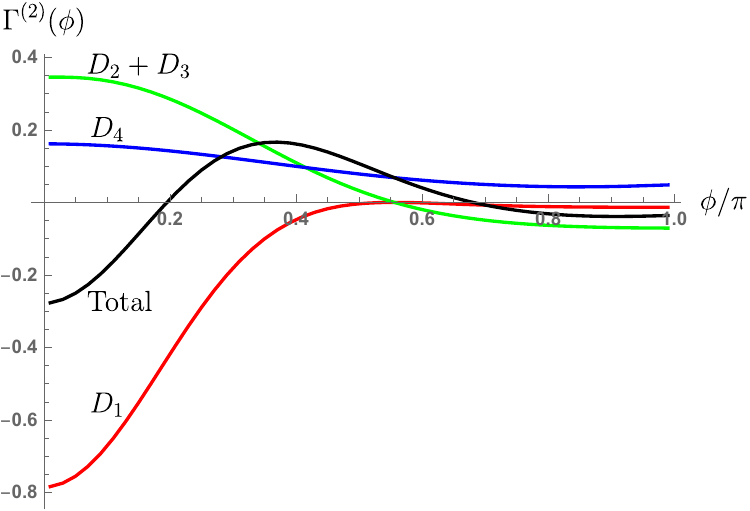}
\caption{The second order contribution to the effective interaction, $\Gamma^{(2)}(\phi)$. It is defined in Eq. \eqref{eq:GM2} and contains the contributions from
 four Kohn-Luttinger diagrams, $D_1$ to $D_4$. Most noticeably, $D_1$ turns negative (attractive).
The black curve is the total sum of all four diagrams. $k_FR=2.5$. }
\label{fig:interaction-2nd-order}
\end{figure}

\begin{figure}
\includegraphics[width=0.45\textwidth]{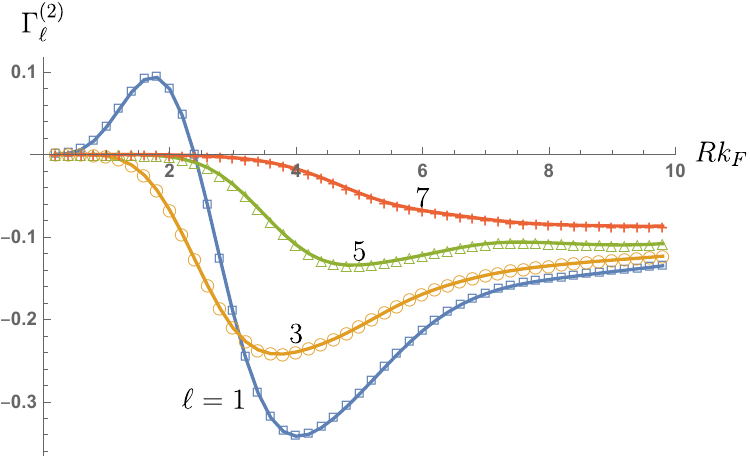}
\caption{The second order contribution to pairing interaction $\Gamma^{(2)}_\ell$ 
in angular momentum channel $\ell=1$, $3$, $5$, and $7$.
They all become attractive for sufficiently large $k_FR$.}
\label{fig:component-2nd-order}
\end{figure}

The second-order contribution to the effective interaction is given by 
summing over $D_1$ to $D_4$ for $\mathbf{p}$ and $\mathbf{p}'$ on the Fermi surface. 
In unit of $(2\pi V_0R^2)$, the result can be organized into 
%a $g$-independent form,
\begin{equation}
\Gamma^{(2)}(\phi) = \frac{\pi}{g}\sum_{i=1}^4 D_i . \label{eq:GM2}
\end{equation}
As an example, the function $\Gamma^{(2)}(\phi)$ for the case of 
$Rk_F=2.5$ is plotted in Fig. \ref{fig:interaction-2nd-order}. We find that 
the contribution for diagram $D_1$ (the red curve) turns negative 
for a significant range of $\phi$ values, e.g. $\phi<\pi$, while
in the same region the contributions from $D_2+D_3$ (in green) and $D_4$ (in blue) 
remain positive. As a result, the total sum (the black curve) develops 
oscillations with $\phi$. 
This clearly shows that density fluctuations as captured by $D_1$ plays an
important role in making the pairing glue.
We can further decompose 
%the second order correction to effective interaction 
$\Gamma^{(2)}(\phi)$ into angular momentum channels, 
the resulting $\Gamma^{(2)}_\ell$ for $\ell=1$ (blue square), $\ell=3$ (orange circle), $\ell=5$ (green triangle), 
and $\ell=7$ (red plus) are shown in Fig. \ref{fig:component-2nd-order}. One observes that as $R$ is increased, all components eventually turn attractive.
%In contrast to the bare value (positive), we now see that the second order correction
%to the effective interaction is attractive for $l=3$ and $l=5$. 
For $\ell=3$, the effect is most pronounced around $R\sim 3.7/k_F$. 
%Note that this is not Kohn-Luttinger physics at large $l$, and also differs from spin fluctuations in the Hubbard model.

\begin{figure}
\includegraphics[width=0.45\textwidth]{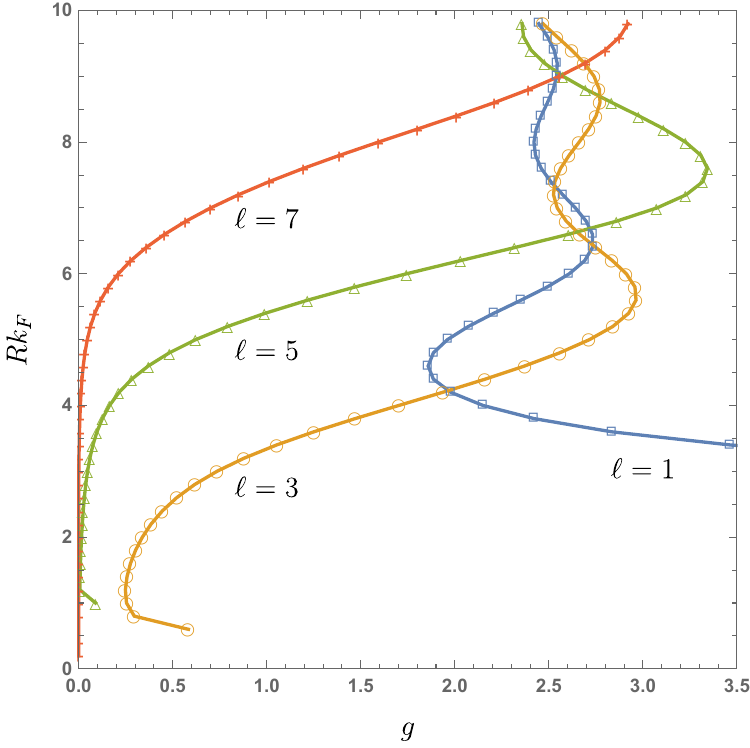}
\caption{The onset of superfluid phases with $\ell=1$ (blue), $3$ (orange), $5$ (green), and $7$ (red) 
according to second order perturbation theory. The data points represent the critical value
$g_c$. For fixed $k_FR$, pairing occurs for $g>g_c$.}
\label{fig:perturbation-boundary}
\end{figure}

Now we can combine the second-order contribution $\Gamma^{(2)}_\ell$ with 
the bare repulsion $\Gamma^{(1)}_\ell$. We ask at what critical values $g=g_c$
the total effective interaction turns attractive, i.e., 
\begin{equation}
\Gamma^{(0)}_\ell + \frac{g_c}{\pi}\Gamma^{(2)}_\ell = 0.
\end{equation}
Solving this equation for $g_c$, we arrive at the perturbative phase diagram in Fig. \ref{fig:perturbation-boundary},
where the phase boundaries of the $\ell=1$, $3$, $5$, $7$ superfluid are plotted using the same symbols as in Fig. \ref{fig:component-2nd-order}. 
The unconventional Cooper pairing discovered here is to some degree parallel to the high partial-wave pairing in the particle-hole (or density wave) channel predicted for fermionic systems with soft-core interactions~\cite{Li2015a}. 
One crucial difference is that the high partial-wave pairing in the particle-particle (superfluid) channel only requires time-reversal or parity symmetries of the Fermi surface whereas the corresponding particle-hole pairing requires Fermi surface nesting effects in addition~\cite{Li2015a}. 

Now we are in position to compare Fig. \ref{fig:perturbation-boundary} to Fig. \ref{fig:phase-diagram}.
According to the second order perturbation theory, $p$-wave pairing (blue square in Fig. \ref{fig:perturbation-boundary})
is pushed toward large $g$ 
and $R$ values. This is mainly because the bare interaction $\Gamma^{(1)}_{\ell=1}$ is large and positive,
and therefore rather hard to overcome. Another reason is that $\Gamma^{(2)}_{\ell=1}$ only becomes
negative when $R>2.4$ as shown in Fig. \ref{fig:component-2nd-order}. Note that according to FRG which contains many more diagrams to higher order, $p$-wave pairing is actually absent from the phase diagram. 
The onset of $f$-wave pairing (orange circles) in Fig. \ref{fig:perturbation-boundary} is roughly consistent with 
the FRG phase boundary except for large $R$. It is stabilized within the window between $k_FR\sim 2$ and $k_FR\sim 4$,
where the bare repulsion $\Gamma^{(1)}_{\ell=3}$ is not particularly strong but $\Gamma^{(2)}_{\ell=3}$ already turns negative.
For these reasons, the onset of $f$-wave superfluid requires much a smaller $g_c$ than the $p$-wave.
In short, perturbation theory correctly predicts that $f$-wave superfluid is preferred over $p$-wave in our model. 
Pairing with larger $\ell$ moves successively to larger $R$ and lower $g_c$,  and the relative positions of the $\ell=3$, $5$, $7$ 
lobes from Fig. \ref{fig:perturbation-boundary} are roughly in line with Fig. \ref{fig:phase-diagram}. 

In summary, it is fair to say that the perturbative calculation above captures some of the
rough features of the FRG phase diagram. On the one hand, it is able to pinpoint certain microscopic processes (e.g. $D_1$ to $D_4$)
that work together to turn the effective interaction attractive, i.e. to provide the pairing glue. On the other hand,
the details of Fig. \ref{fig:perturbation-boundary} differ significantly from Fig. \ref{fig:phase-diagram}.
This is not surprising, for the perturbation results are not reliable at higher $k_FR$ values.

\section{Summary and Outlook}
%\section{ {\red Summary and Outlook} } 
We have presented evidence for superfluid phases with Cooper pair angular momentum $\ell=3$, $5$, $7$, $9$ in 
a model system of spin-polarized fermions with short-range repulsive interactions. Our main goal is to elucidate
how the repulsion is turned into glue that binds the fermions into Cooper pairs. While FRG provides the full picture and 
more accurate results, some of the trends and gross features can already be appreciated from perturbative considerations. 
 {\blue According to our calculation, it is inaccurate to only} credit density fluctuations such as diagram $D_1$ for providing the glue, 
because other processes also contribute to the renormalization of the effective interaction, e.g. to the second order correction $\Gamma^{(2)}_\ell$.
{\blue Comparing the phase diagram Fig. 1 with the case of Rydberg-dressed Fermi gas \cite{PhysRevA.101.023624} 
clearly shows that {\it the form of the bare interaction matters}}.

These considerations naturally lead to the open question: 
assuming that we can engineer arbitrary $v(r)$ using the tricks of Atomic Molecular and Optical physics,
which kind of bare repulsive interaction $v(r)$ offers the best route toward superfluid with a reasonably high $T_c$? 
A heuristic argument is that we would like
$v(r)$ to have sharp features, so that its Fourier transform $v({q})$ will acquire negative segments which could be potentially advantageous
to pairing. While this intuition serves us well by inspiring the choice of Eq. \eqref{disk} in the present work,
it must be kept in mind that this is not a first order effect. For example, in our example, to the first order of $V_0$, 
all $\Gamma^{(1)}_\ell>0$; one must carefully compute 
the effective interaction by taking many-body processes into account. 
{\blue Roughly speaking, higher angular momentum (rather than $p$-wave) pairing is preferred because
there is less bare repulsion to overcome,
and it can take better advantage of the oscillation of $\Gamma^{(2)}(\phi)$ around the Fermi surface. 
It may be challenging to realize the simple model and the phases predicted here in near future experiments.
But the lessons learned from the case study, including the general trend and the underlying mechanism, 
can benefit the ongoing effort to engineer stronger pair glue in repulsive Fermi gases.}

%In a broader perspective, recently soft-core interaction is proposed for spin squeezing %\cite{PhysRevResearch.5.L012033}. 
Previously 
$f$-wave pairing has been discussed for example in the context of superfluid helium three \cite{PhysRevB.34.4861,
PhysRevLett.97.115301} as well as cold atoms on optical lattice \cite{PhysRevA.82.053611},
but in those cases it is stabilized by very different mechanisms. 
We stress that in the present work, the bare interaction is repulsive and the system is two dimensional. 
This differs from previous studies on 
Rydberg-dressed Fermi gas in 3D with attractive interactions \cite{PhysRevA.90.013631}
including the appearance of high partial wave pairing \cite{PhysRevA.104.L061302} by coupling to a $nD$ state.
In our case, pairing beyond $f$-wave (with $\ell\geq 5$) requires larger value of $k_FR$, i.e. away from the dilute limit. 
{\blue It remains an open problem regarding what happens if we generalize the model to spin-$1/2$ Fermi gases, where the effect of  
long-range potentials on pairing has been discussed \cite{kabanov2011,kivelson2012}. 
Whether the $f$-wave pairing found here can lead to topological superfluid state is another 
question left for future study.}
%whether fluctuations give rise to non-Fermi liquid behaviors at finite temperatures. 

\begin{acknowledgements}
This work is supported by NSF grant PHY-206419 and AFOSR grant FA9550-23-1-0598 (EZ), TUBITAK
2236 Co-funded Brain Circulation Scheme 2 (CoCirculation2) Project No. 120C066 (AK), 
National Program on Key Basic Research Project of China Grant 2021YFA1400900 (XL), 
and National Natural Science Foundation of China Grant 11934002 (XL).
EZ acknowledges illuminating discussions with A. Chubukov, and he
is indebted to J. A. Sauls for questions that led to the comparison between $p$- and $f$-wave pairing in Section IV. 
Part of this work (EZ) is performed at 
Aspen Center for Physics, which is supported by NSF grant PHY-2210452.
\end{acknowledgements}
\bibliography{papers}
\end{document}